\documentstyle[aps]{revtex}
\begin{document}
\voffset 0.8truecm
\title{Phase-covariant quantum cloning of qudits
}

\author{
Heng Fan, Hiroshi Imai, Keiji Matsumoto,
Xiang-Bin Wang
}
\address{
Quantum computation and quantum information project,
ERATO, \\
Japan Science and Technology Corporation,\\
Daini Hongo White Building 201,
Hongo 5-28-3, Bunkyo-ku, Tokyo 133-0033, Japan.
}
\maketitle

\begin{abstract}
We study the phase-covariant quantum cloning machine for qudits,
i.e. the input states in d-level quantum system
 have complex
coefficients with arbitrary phase but constant module.  A cloning
unitary transformation is proposed. After optimizing the fidelity
between input state and single qudit reduced density opertor of output
state, we obtain the optimal fidelity for
1 to 2 phase-covariant quantum
cloning of qudits and the corresponding cloning transformation.
\end{abstract}


\section{Introduction}
No-cloning theorem \cite{WZ} is one of the most fundamental differences
between classical and quantum information. It states an
arbitrary quantum state cannot be cloned exactly.
No-cloning theorem is also extended to other cases such as
no broadcasting\cite{BCFJ,KI}, no-imprinting\cite{M}.
A unified principle was proposed recently\cite{KI1}.
While an arbitrary quantum state cannot be cloned perfectly,
we can clone it approximately\cite{BH} or probabilisticaly\cite{DG}.
Thus some quantum cloning machines are proposed to
study the cloning of quantum states.

A universal quantum cloning machine (UQCM) proposed
by Bu\v{z}ek and Hillery\cite{BH} clones an arbitrary
quantum state approximately. The quality of the
copies is independent of the input state. This universal
quantum cloning machine is studied and is generalized in several directions.
Using fidelity as the measurement of quality of copies,
Bu\v{z}ek and Hillery's cloning machine is proved to be optimal
\cite{BDEF}. Instead of a single input qubit and
two copies, the UQCM
with general $N$ identical pure input qubits and $M$ copies was
studied in \cite{GM}, the optimal fidelity is also obtained.
By identifying the fidelity of copying
$N$ identical qubits to infinite copies with
the fidelity of the corresponding quantum state estimation\cite{DBE},
the upper bound of fidelity of UQCM can be found\cite{BEM}.
Besides the cloning of qubits, the UQCM for d-level
quantum states, qudits, is studied by completely positive map\cite{W,Z}.
The unitary transformation for cloning of qudits was
studied in \cite{BH1} for 1 to 2 case, in \cite{FMW} for general
$N$ to $M$ case. The physical implementation of universal
quantum cloning machine was proposed in \cite{SWZ,KSW}.
Quantum networks to realize quantum cloning machine was
studied in \cite{BBHB}.

A UQCM copies arbitrary pure quantum states equally well.  So, we can
use UQCM in the case that the input state is completely unknown.
However, sometimes, we already know partial information about the
input state.
 If we know exactly the input quantum
state, we can clone it perfectly.  If we do not know it exactly, but have
partial information about it, we can perhaps design a special
quantum cloning machine for this kind of input state with a better quality
than the UQCM.
A  phase-covariant quantum cloning is such a special quantum
cloning machine.  It is defined as a machine that optimally clone
a special class of states,
the states that have complex coefficients
with arbitrary phase but constant module (see (\ref{input})).  In
2-level quantum system, this special class of states is one kind of
equatorial qubits. Here equatorial qubit means that one parameter of
its Bloch vector is zero. We can change phase-covariant quantum
cloning machine to the cloning machine for other equatorial qubits
input via some unitary transformations. So, we generally do not
distinguish phase-covariant quantum cloning machine with cloning
machine for equatorial qubits.  For 2-level quantum system, the 1 to 2
phase-covariant quantum cloning machine was studied in \cite{BCDM}.
The 1 to M cloning machine for equatorial qubits was studied in
\cite{FMWW} and the fidelity was proved to be optimal.  The
phase-covariant quantum cloning machine is of interest in particular
in quantum key distribution.  In the optimal eavesdropping of
BB84\cite{BB84} quantum key distribution, instead of a UQCM, the
eavesdropper should use the phase-covariant quantum cloning machine
instead of the well studied UQCM\cite{BCDM}. If all 3 mutually
unbiased states in 2-level system are used, i.e. the 6-state quantum
key distribution scheme, a UQCM should be used\cite{bruss}.  Besides
the 2-level quantum system, the phase-covariant quantum cloning
machine in 3-level quantum system is also studied. By different
methods, the optimal phase-covariant quantum cloning machine for
3-level quantum system is obtained by two groups\cite{DP,CDG}.

In this paper, we shall study the phase-covariant quantum cloning
machine in d-level quantum system.
We assume that the output states of the phase-covariant
cloning machine are symmetric as in UQCM \cite{BH,GM,W}.
We find a simple unitary transformation for our d-level
phase-covariant quantum cloning machine. Next, we optimize the
fidelity over free parameters.  As expected, the optimal fidelity for
qudit is higher than the corresponding UQCM.  For special case, $d=2$
and $d=3$, the optimal fidelity obtained in this paper agree with
previous known results.  For case $d$ is a prime number, we point out
this optimal phase-covariant quantum cloning machine can be used in
eavesdropping of quantum key distribution by using $d$ mutually
unbiased states.  However, it is not neccessarily optimal for
eavesdropping since optimal cloning is not known to be equivalent to
optimal eavesdropping in general.

\section{Some known results about phase-covariant quantum cloning machine} 
We first introduce the notations and review some known results for
qubits\cite{BCDM,FMWW}.  We consider the input state as
\begin{eqnarray}
|\Psi \rangle ^{(in)}&=&\frac {1}{\sqrt{2}}[|0\rangle +e^{i\phi }
|1\rangle ],
\label{2input}
\end{eqnarray}
where $\phi \in [0,2\pi )$. This state just has one arbitrary phase
parameter $\phi $ instead of two free parameters for an arbitrary
qubit. So, we already know partial information of this input state.
One can check that the $y$ component of the Bloch vector of this state
is zero.  This case is equivalent to the case that the input state is
$|\Psi \rangle =\cos \theta |0\rangle +\sin \theta |1\rangle $, in
which the input state does not have arbitrary phase parameter.  The
optimal phase-covariant cloning transformation takes the form,
\begin{eqnarray}
U|0\rangle ^{(in)}|Q\rangle =
\frac {1}{\sqrt{2}}|00\rangle |0\rangle _a
+{\frac 12}\left( |01\rangle +
|10\rangle \right) |1\rangle _a,
\nonumber \\
U|1\rangle ^{(in)}|Q\rangle =
\frac {1}{\sqrt{2}}|11\rangle |1\rangle _a
+{\frac 12}\left( |01\rangle +
|10\rangle \right) |0\rangle _a,
\label{2dclone}
\end{eqnarray}
where $|Q\rangle $ is the blank state and initial state of the cloning
machine. The first states in l.h.s. are input states.  The states with
subindices $a$ are ancilla states of cloning machine which should be
traced out to obtain the output state.  The copies appear in the first
two qubits in r.h.s., actually the first two qubits are symmetric so
that the reduced density matrices of copies are equal.  The single
qubit reduced density matrix of output can be calculated as
\begin{eqnarray}
\rho ^{out}_{red.}=\frac {1}{\sqrt{2}}\rho ^{(in)}+
\left( {1\over 2}-\sqrt{1\over 8}\right) I,
\label{scalar}
\end{eqnarray}
where $I$ is the identity matrix, and the input density matrix is
$\rho ^{(in)}=|\Psi \rangle \langle \Psi |$ defined in
(\ref{2input}). We use fidelity to define the quality of the
copies. The general definition of fidelity takes the form $F(\rho _1,
\rho _2)=[Tr\sqrt {(\rho _1^{1/2}\rho _2\rho _1^{1/2})}]^2$
\cite{Jozsa}.  The value of $F$ ranges from 0 to 1. A larger F
corresponds to a higher fidelity. $F=1$ means two density matrices are
equal.  We only consider about the pure input states, and the fidelity
can be simplified as $F=^{(in)}\langle \Psi |\rho ^{(out)}_{red.}|\Psi
\rangle ^{(in)}$.  The optimal fidelity of phase-covariant quantum
cloning machine is obtained as
\begin{eqnarray}
F_{optimal}={1\over 2}+\sqrt{1\over 8}.
\label{level2}
\end{eqnarray}
As expected, this fidelity $F\approx 0.85$ is higher than
the fidelity of UQCM $F\approx 0.83$.

For the well known BB84 quantum key distribution protocol, 
all four states $|0\rangle ,|1\rangle , 1/\sqrt{2}(|0\rangle
+|1\rangle ), 1/\sqrt{2}(|0\rangle -|1\rangle )$ can be written
as the form $|\Psi \rangle =\cos \theta |0\rangle +\sin \theta |1\rangle $.
So, instead of the UQCM, we should at least use the cloning machine for
equatorial qubits in eavesdropping.  Actually in individual attack, we
can not do better than the cloning machine for equatorial
qubits\cite{BCDM,FGGNP}. The cloning machine presented in
(\ref{2dclone}) can be used to analyze
the eavesdropping if other 2
mutually unbiased bases $1/\sqrt{2}(|0\rangle -|1\rangle ),
1/\sqrt{2}(|0\rangle +|1\rangle ), 1/\sqrt{2}(|0\rangle +i|1\rangle ),
1/\sqrt{2}(|0\rangle -i|1\rangle )$ are used.

The optimal fidelity of phase-covariant quantum cloning machine for
qutrits ($d=3$) was obtained by D'Ariano et al\cite{DP} and Cerf et
al\cite{CDG};
\begin{eqnarray}
F=\frac {5+\sqrt{17}}{12}.
\label{level3}
\end{eqnarray}

\section{Phase-covariant cloning of qudits}
We study the quantum cloning of d-level states in the form
\begin{eqnarray}
|\Psi \rangle ^{(in)}=\frac {1}{\sqrt{d}}\sum _{j=0}^{d-1}
e^{i\phi _j}|j\rangle ,
\label{input}
\end{eqnarray}
where $\phi _j\in [0,2\pi ), j=0,
\cdots, d-1$, are arbitrary phase paramters.
A whole phase is not important, so we can assume $\phi
_0=0$.  The density operator of input state can be written as $\rho
^{(in)}={1\over d}\sum _{jj'}e^{i(\phi _j-\phi _{j'})}|j\rangle \langle
j'|$.  For case 1 to M phase-covariant quantum cloning machine (with 1
input qudit and M output qudits), we will use the assumption that the
most general cloning transformation takes the following form
\begin{eqnarray}
U|j\rangle |Q\rangle =\sum _{\vec{k}}^M|\vec{k}\rangle
|R_{j\vec{k}}\rangle ,
\label{general}
\end{eqnarray}
where similar notations as in 2-level quantum system are used, and
$\vec{k}\equiv \{ k_0, \cdots, k_{d-1}\} $,
the quantum state
$|\vec{k}\rangle $ is a normalized symmetric state with $k_j$ states
in $|j\rangle $.
The summation $\sum
_{\vec{k}}^M$ means taking sum over all possible values satisfy
the restriction $\sum _{j=0}^{d-1}k_j=M$.
The ancilla states $|R_{j\vec{k}}\rangle $ are not
necessarily orthogonal and normalized.
The unitarity of the cloning transformation (\ref{general}) means the
restriction $\sum _{\vec{k}}^M\langle
R_{j\vec{k}}|R_{j'\vec{k}}\rangle =\delta _{jj'}$.  We remark here
that as in UQCM, the output states are symmtrical so that all single
qudit reduced density matrix of output are equal to each other.
Except the assumption that the output states are symmetric as in UQCM
\cite{BH,GM,W}, the relation (\ref{general}) is the most general
cloning transformation.
Using this assumption, we can find the optimal phase-covariant cloning
machine from (\ref{general}).  Substituting the input state
(\ref{input}) into the general cloning machine (\ref{general}),
tracing out the ancilla states, we have the output state as follows,
\begin{eqnarray}
\rho ^{(out)}=\frac {1}{d}
\sum _{j,j'=0}^{d-1}\sum _{\vec{k},\vec{k'}}^M
e^{i(\phi _j-\phi _{j'})}|\vec{k}\rangle \langle \vec{k'}|
\langle R_{j'\vec{k'}}|R_{j\vec{k}}\rangle .
\label{wholeout}
\end{eqnarray}
Now, let us calculate the single qudit
reduced density operator of the output.
The diagonal elements of the output reduced density operator
can derive from the term 
$|\vec{k}\rangle \langle \vec{k}|$. The off-diagonal elements
$|l\rangle \langle l'|$ of the output reduced density operator
can derive from the term
$|\vec{k}\rangle \langle \vec{k'}|$ in case ${\vec {k}}$ and
${\vec {k'}}$ satisfy $k_l=k_l'+1, k_{l'}+1=k'_{l'}$, and other
elements are equal. We can find the reduced density
operator of output takes the following form,
\begin{eqnarray}    
\rho _{red.}^{(out)}&=&\sum _{l=0}^{d-1}|l\rangle \langle l|
\left[ {1\over d}\sum _{j,j'=0}^{d-1}e^{i(\phi _j-\phi _{j'})}
\sum _{\vec{k}}^M
\frac {k_l}{M}\langle R_{j'\vec{k}}|R_{j\vec{k}}\rangle \right]
\nonumber \\
&&+\sum _{l\not= l'}|l\rangle \langle l'|
\left[ {1\over d}
\sum _{j,j'=0}^{d-1}e^{i(\phi _j-\phi _{j'})}
\sum _{\vec{k}\vec{k'}}^M\frac {\sqrt{k_lk'_{l'}}}M
\langle R_{j'\vec{k'}}|R_{j\vec{k}}\rangle
\delta _{k_0,k'_0}
\cdots \delta _{k_{d-1},k'_{d-1}}
\delta _{k_l,k'_l+1}\delta _{k_{l'}+1, k'_{l'}}
\right] ,
\label{gout1}
\end{eqnarray}
where in $\cdots $, we do not have $\delta _{k_l,k'_l}$ and
$\delta _{k_{l'},k'_{l'}}$, the same notations will be used later.
Due to the output reduced density operator (\ref{gout1}), 
the request for phase covariance implies the following
restriction
\begin{eqnarray}
&&\sum _{\vec{k}}^M
\frac {k_l}{M}\langle R_{j'\vec{k}}|R_{j\vec{k}}\rangle
\propto \delta _{jj'},
\label{orth} \\
&&\sum _{\vec{k}\vec{k'}}^M\frac {\sqrt{k_lk'_{l'}}}M
\langle R_{j'\vec{k'}}|R_{j\vec{k}}\rangle
\propto \delta _{jl}\delta _{j'l'},
\nonumber \\
&&~~~~~~{\rm if}~ k_l=k_l'+1, k_{l'}+1=k_{l'}',~ k_m=k'_m,~ m\not=l,l'.
\label{simple}
\end{eqnarray}
So, the output single qudit reduced density matrix can be written as
the following form
\begin{eqnarray}
\rho _{red.}^{(out)}&=&\sum _{l=0}^{d-1}|l\rangle \langle l|
\left[ {1\over d}\sum _{j=0}^{d-1}\sum _{\vec{k}}^M
\frac {k_l}{M}\langle R_{j\vec{k}}|R_{j\vec{k}}\rangle \right]
\nonumber \\
&&+\sum _{j\not= j'}e^{i(\phi _j-\phi _{j'})}|j\rangle \langle j'|
\left[ {1\over d}\sum _{\vec{k}\vec{k'}}^M\frac {\sqrt{k_jk'_{j'}}}M
\langle R_{j'\vec{k'}}|R_{j\vec{k}}\rangle
\delta _{k_0,k'_0}
\cdots \delta _{k_{d-1},k'_{d-1}}
\delta _{k_j,k'_j+1}\delta _{k_{j'}+1, k'_{j'}}
\right] .
\label{gout}
\end{eqnarray}
The corresponding fidelity is written as
\begin{eqnarray}
F={1\over d}+{1\over {d^2}}
\left[ \sum_{j\not= j'}\sum _{\vec{k}\vec{k'}}^M\frac {\sqrt{k_jk'_{j'}}}M
\langle R_{j'\vec{k'}}|R_{j\vec{k}}\rangle
\delta _{k_0,k'_0}
\cdots \delta _{k_{d-1},k'_{d-1}}
\delta _{k_j,k'_j+1}\delta _{k_{j'}+1, k'_{j'}}
\right] .
\label{gfidelity}
\end{eqnarray}

Next, we shall pay our attention to 1 to 2 phase-covariant
quantum cloning machine.
Considering the restriction
(\ref{orth},\ref{simple}), and also considering the
symmetric property of the input state (\ref{input}),
we have the following phase-covariant quantum cloning transformation
\begin{eqnarray}
U|j\rangle |Q\rangle =\alpha |jj\rangle |R_j\rangle
+\frac {\beta }{\sqrt{2(d-1)}}\sum _{l\not =j}^{d-1}
(|jl\rangle +|lj\rangle )|R_l\rangle ,
\label{clone}
\end{eqnarray}
where $\alpha ,\beta $ are real numbers, and $\alpha ^2+\beta ^2=1$.
Actually letting $\alpha , \beta $ to be complex numbers does not
improve the fidelity. $|R_j\rangle $ are orthonormal ancilla states.
This is a simplified cloning transformation.
In the most general cloning transformation (\ref{general}), the
ancilla states should be denoted as $|R_{j\vec{k}}\rangle $.
In case 1 to 2 cloning, there are four kinds of ancilla states,
$|R_{j,k_j=2}\rangle $,$|R_{j,k_j=1,k_l=1}\rangle $,
$|R_{j,k_l=2}\rangle $, $|R_{j,k_l=1,k_m=1}\rangle $, where
$j\not= l\not= m$. However, we can set the last two ancilla
states as zeroes without affecting any kinds of restrictions
and loosing fidelity.
And also
the ancilla state $|R_{j'\vec{k'}}\rangle $
can be identified with $|R_{j\vec{k}}\rangle $ when
$k'_j=1, k'_{j'}=1, k_j=2$ with some normalization
due to (\ref{orth},\ref{simple}).
That means we can take
$|R_{j,k_j=2}\rangle \propto |R_{j,k_j=1,k_l=1}\rangle $.
So, we actually
just need one ancilla state $|R_j\rangle $ to represent
$|R_{j\vec{k}}\rangle $ and $|R_{j'\vec{k'}}\rangle $
if we have relations $k_j=2, k_{j'}=0;  k'_j=1, k'_{j'}=1$.
Without other states in (\ref{clone}),  the cloning transformation
(\ref{clone}) can achieve the optimal fidelity due to
relation (\ref{gfidelity}). In short, we can
find the optimal cloning transformation from (\ref{clone}).

Substituting the input state (\ref{input}) into the cloning transformation
and tracing out the ancilla states, the output state takes the form
\begin{eqnarray}
\rho ^{(out)}&=&\frac {\alpha ^2}{d}\sum _j|jj\rangle \langle jj|
+\frac {\alpha \beta }{d\sqrt {2(d-1)}}\sum _{j\not =l}
e^{i(\phi _j-\phi _l)}\left[ |jj\rangle (\langle jl|+\langle |lj|)
+(|jl\rangle +|lj\rangle )\langle ll|\right]
\\
&&+\frac {\beta ^2}{2d(d-1)}\sum _{jj'}\sum _{l\not =j,j'}
e^{i(\phi _j-\phi _{j'})}(|jl\rangle +|lj\rangle )
(\langle lj'|+\langle j'l|).
\end{eqnarray}
Taking trace over one qudit, we obtain the single qudit
reduced density matrix of output
\begin{eqnarray}
\rho ^{(out)}_{red.}={1\over d}\sum _{j}|j\rangle \langle j|
+\left( \frac {\alpha \beta }{d}\sqrt{\frac {2}{d-1}}
+\frac {\beta ^2(d-2)}{2d(d-1)}\right) \sum _{j\not =k}
e^{i(\phi _j-\phi _k)}|j\rangle \langle k|.
\label{finalscalar}
\end{eqnarray}
The fidelity can be calculated as
\begin{eqnarray}
F={1\over d}+\alpha \beta \frac {\sqrt{2(d-1)}}{d}
+\beta ^2\frac {d-2}{2d}.
\end{eqnarray}
These relations are the special cases of the general 1 to M
cloning that we obtained in (\ref{gout},\ref{gfidelity}), but with $M=2$.
Now, we need to optimize the fidelity under the restriction
$\alpha ^2+\beta ^2=1$. We can find the optimal fidelity
of 1 to 2 phase-covariant quantum cloning machine as
\begin{eqnarray}
F_{optimal}=\frac {1}{d}+\frac {1}{4d}(d-2+\sqrt{d^2+4d-4}).
\label{dfidelity}
\end{eqnarray}
In cases $d=2,3$, these results agree with the previous known results
(\ref{level2},\ref{level3}), respectively.
As expeceted, the optimal fidelity of phase-covariant
quantum cloning machine is higher than the
corresponding optimal fidelity of UQCM $F_{optimal}>
F_{universal}=(d+3)/2(d+1)$.
The optimal fidelity can be achieved when $\alpha ,\beta $ take
the following values,
\begin{eqnarray}
\alpha =\left( {1\over 2}-\frac {d-2}{2\sqrt{d^2+4d-4}}\right)
^{1 \over 2},\nonumber \\
\beta =\left( {1\over 2}+\frac {d-2}{2\sqrt{d^2+4d-4}}\right)
^{1 \over 2}.
\label{ab}
\end{eqnarray}
In case $d=2$, the cloning transformation (\ref{clone}) recovers
the previous result (\ref{2dclone}).

Thus we find the optimal 1 to 2 phase-covariant quantum  cloning
machine for qudits (\ref{clone}, \ref{ab}) and
the corresponding optimal fidelity (\ref{dfidelity}).
We remark that the output reduced density operator
in (\ref{finalscalar}) can be written as the scalar form \cite{BM}
with respect to the input density operator
$\rho
^{(in)}={1\over d}\sum _{jk}e^{i(\phi _j-\phi _{j'})}|j\rangle \langle
j'|$ as in 2-level system (\ref{scalar}).

\section{Discussions and summary}
Quantum measurements by mutually unbiased bases provide the
optimal way of determining a quantum state. And the
mutually unbiased bases have close relations with quantum
cryptography. In d-dimension, when d is prime, there
are $d+1$ mutually unbiased bases. Except the standard basis
$\{ |0\rangle, |1\rangle , \cdots, |d-1\rangle \} $, the
other $d$ mutually unbiased bases take the form \cite{BBRV}
\begin{eqnarray}
|\psi _t^l\rangle =\frac {1}{\sqrt{d}}
\sum _{j=0}^{d-1}(\omega ^t)^{d-j}(\omega ^{-k})^{s_j}|j\rangle ,
~~~t=0, \cdots, d-1,
\label{mub}
\end{eqnarray}
where $s_j=j+\cdots +(d-1)$. And $l=0, \cdots, d-1$ represent
$d$ mutually unbiased bases. The phase-covariant quantum cloning
machine of qudits can clone all of these states equally well.
So, we see if one uses $d$ mutually unbiased bases (\ref{mub}) to perform
quantum key distribution, the eavesdropper could use
phase-covariant quantum cloning machine to attack instead of
the UQCM. If all $d+1$ mutually unbiased bases are used,
we should use UQCM.
However, it is not known
whether using phase-covariant cloning machine in eavesdropping
is optimal or not when $d~(d>3)$ bases are used
even though the cloning machine itself
is optimal.
The difference between quantum key distribution
schemes by using $d$ and
$d+1$ mutually unbiased bases decreases when $d$ becomes larger.
Correspondingly the gap between the fidelities of phase-covariant
cloning machine and UQCM decreases when $d$ becomes larger.
When $d$ is large enough, this gap becomes negligible.

In summary, we present in this paper the optimal
1 to 2 phase-covariant quantum cloning machine for qudits
(\ref{clone}, \ref{ab}).
The corresponding optimal fidelity (\ref{dfidelity}) was found.
In $d=2$ case, the results recover the previous results \cite{BCDM,FMW}.
In $d=3$, the optimal fidelity  agree with the result obtained
by D'Ariano et al\cite{DP} and Cerf et al \cite{CDG}.

{\it Acknowlegements}: We would like to thank Dominic Mayers
for a lot of very useful suggestions and revisions of this
paper which improved substantially this paper.
We also thank the anonymous referee for pointing out some
mistakes in the first version of this paper.
We thank Miki Wadati for critical reading this manuscript.

\end{document}